%% file: main.tex
\documentclass[conference]{IEEEtran}
\IEEEoverridecommandlockouts

\usepackage{cite}
\usepackage{amsmath,amssymb,amsfonts}
\usepackage{algorithmic}
\usepackage{algorithm}
\usepackage{graphicx}
\usepackage{textcomp}
\usepackage{xcolor}
\usepackage{hyperref}
\usepackage{booktabs}
\usepackage{multirow}
\usepackage{subcaption}
\usepackage{tikz}
\usepackage{pgfplots}
\definecolor{myPink}{HTML}{FAD9D5}
\usepackage[table]{xcolor}   
\pgfplotsset{compat=1.17}
\usepackage{enumitem}
\def\BibTeX{{\rm B\kern-.05em{\sc i\kern-.025em b}\kern-.08em
    T\kern-.1667em\lower.7ex\hbox{E}\kern-.125emX}}
\newcommand{\mypara}[1]{\noindent\textbf{#1}\quad}

\begin{document}

\title{Readout-Side Bypass for Residual Hybrid Quantum-Classical Models}

\author{
Guilin Zhang\IEEEauthorrefmark{1},
Wulan Guo\IEEEauthorrefmark{1},
Ziqi Tan\IEEEauthorrefmark{1},
Hongyang He\IEEEauthorrefmark{2},
Qiang Guan\IEEEauthorrefmark{3}
Hailong Jiang\IEEEauthorrefmark{4}\IEEEauthorrefmark{5}\thanks{\IEEEauthorrefmark{5} Corresponding author: Hailong Jiang, hjiang@ysu.edu}\\
\IEEEauthorblockA{\IEEEauthorrefmark{1}Department of Engineering Management and Systems Engineering, George Washington University, USA}
\IEEEauthorblockA{\IEEEauthorrefmark{2}Department of Computer Science, University of Warwick, UK}
\IEEEauthorblockA{\IEEEauthorrefmark{3}Department Computer Science, Kent State University, USA}
\IEEEauthorblockA{\IEEEauthorrefmark{4}Department of Computer Science, Information, and Engineering Technology, Youngstown State University, USA\\
\IEEEauthorrefmark{1}\{guilin.zhang, wulan.guo, ziqi.tan\}@gwu.edu,  \IEEEauthorrefmark{2}hongyang.he@warwick.ac.uk,   \IEEEauthorrefmark{3}qguan@kent.edu, \IEEEauthorrefmark{4}hjiang@ysu.edu}
}

\maketitle

\input{sections/abstract}
\input{sections/introduction}
\input{sections/methodology}

\input{sections/experiments}

\input{sections/results}
\input{sections/conclusion}

\clearpage
\bibliographystyle{IEEEtran}
\bibliography{references}

\end{document}

%% file: sections/abstract.tex
\begin{abstract}
Quantum machine learning (QML) promises compact and expressive representations, but suffers from the \emph{measurement bottleneck}---a narrow quantum-to-classical readout that limits performance and amplifies privacy risk. We propose a lightweight \emph{residual hybrid architecture} that concatenates quantum features with raw inputs before classification, bypassing the bottleneck without increasing quantum complexity. Experiments show our model outperforms pure quantum and prior hybrid models in both centralized and federated settings. It achieves up to \textbf{+55\% accuracy improvement} over quantum baselines, while retaining \textbf{low communication cost} and \textbf{enhanced privacy robustness}. Ablation studies confirm the effectiveness of the residual connection at the quantum-classical interface. Our method offers a practical, near-term pathway for integrating quantum models into privacy-sensitive, resource-constrained settings like federated edge learning.
\end{abstract}


%% file: sections/introduction.tex
\section{Introduction}
\label{sec:introduction}

Quantum machine learning (QML) offers promising advantages, including compact representations in high-dimensional Hilbert spaces and intrinsic support for nonlinear feature extraction via parameterized quantum circuits (PQCs)~\cite{li2025quantum,khurana2024quantum,xu2025quantum,zaman2023survey}. However, practical QML faces a core limitation: the \emph{measurement bottleneck}, where high-dimensional classical inputs are compressed into a small number of quantum observables. This severely constrains downstream accuracy~\cite{tomar2025comprehensive,rohe2025quantum}.

To mitigate this, \emph{hybrid quantum–classical models} have been proposed, wherein classical neural networks process measured quantum features \cite{havlivcek2019supervised,liu2021hybrid,de2022survey}. Yet, most hybrids remain bottlenecked—only the quantum features $Q(x)$ are passed forward, discarding original inputs $x$ ~\cite{schuld2020circuit}. This narrow interface not only reduces performance but also amplifies susceptibility to \emph{membership inference attacks} (MIA), as compressed features tend to overfit and leak more information ~\cite{farokhi2024maximal}.

In classical deep learning, residual connections help preserve input information across layers and stabilize training~\cite{gomez2017reversible,zhang2019fixup}. However, such residuals bypass internal layers—not the terminal bottleneck. In QML, the information loss occurs at the quantum measurement stage, and traditional residuals are insufficient to address it.

Prior efforts to expand $Q(x)$—via deeper PQCs, multi-basis measurements, or learned readouts~\cite{campbell2024deeperpqc,tasar2025multibasis,chatterjee2025readout}—reshape the feature space but do not expose the original input to the classifier. Moreover, matched-parameter comparisons and privacy evaluations are rarely reported, making it unclear whether such methods are accurate, compact, and robust.

We propose a simple yet effective \emph{readout-side residual hybrid} architecture: concatenate the raw input $x \in \mathbb{R}^d$ with the measured quantum features $Q(x) \in \mathbb{R}^{m'}$ to form $z = [x \| Q(x)]$, then apply a projection $W_{\text{proj}}$ to obtain latent representation $\tilde z$. This non-invasive bypass preserves original inputs, improves robustness, and integrates seamlessly with both centralized and federated settings.

To evaluate the downstream implications of this design, we analyze its privacy robustness under common threat models. We consider a model-release adversary with full access to trained weights, performing MIA via threshold or shadow-model attacks~\cite{geiping2020inverting}. Reconstruction performance is measured by AUC; empirical robustness corresponds to AUC $\approx 0.5$, indicating indistinguishability.

On four benchmarks (Wine~\cite{wine1992}, Breast Cancer~\cite{breastcancer1995}, {Fashion-MNIST Subset}~\cite{xiao2017fashion},
and {Forest CoverType Subset}~\cite{blackard1999comparative}), our residual hybrid achieves near-classical accuracy with 10–20\% fewer parameters and significantly better privacy robustness (MIA AUC $\approx$ 0.5), outperforming both quantum-only and standard hybrid baselines. Ablation confirms that exposing both $x$ and $Q(x)$ to the classifier is crucial for overcoming the measurement bottleneck.

\noindent\textbf{The contributions of this paper are:}
\begin{itemize}
  \item We introduce a \textbf{readout-side residual hybrid} architecture that bypasses the quantum measurement bottleneck.
  \item We conduct \textbf{matched-parameter evaluations} under both accuracy and privacy metrics (MIA AUC) in a model-release threat setting.
  \item We validate on \textbf{real-world datasets} with ablation studies confirming the effectiveness of the bypass mechanism.
\end{itemize}

%% file: sections/methodology.tex
\section{Methodology}
\label{sec:methodology}

\begin{figure*}
    \centering
    \includegraphics[width=0.9\linewidth]{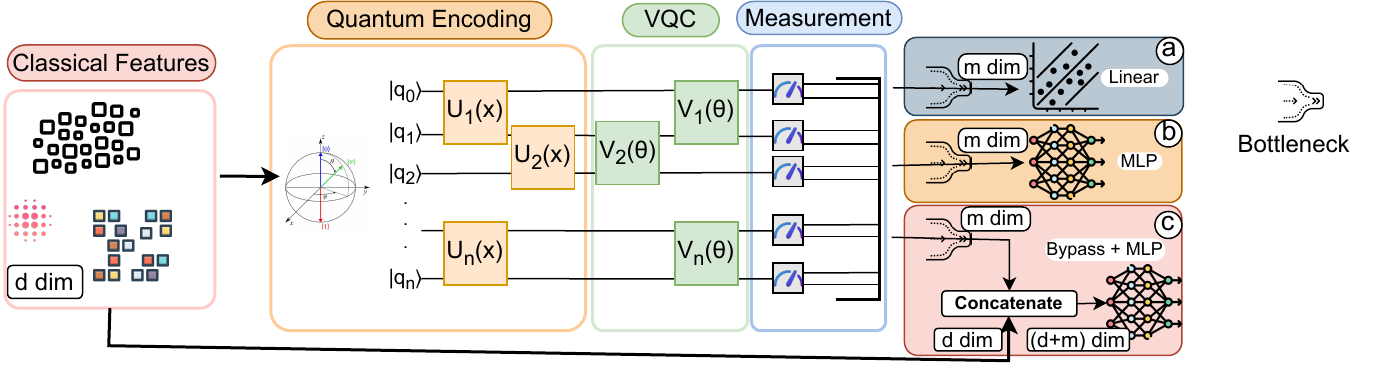}
    \caption{Model comparison. (a) Pure QML: quantum-only readout; (b) Hybrid QML: classical MLP on $Q(x)$; (c) Ours: residual hybrid with $[x \| Q(x)]$ bypassing the measurement bottleneck.}
\label{fig:architecture_comparison}
\end{figure*}

Our method builds upon a shared quantum processing pipeline, shown in Fig. ~\ref{fig:architecture_comparison}. All models encode classical input $x \in \mathbb{R}^d$ into quantum states, transform them via a parameterized quantum circuit (PQC), and measure observables to obtain quantum features $Q(x) \in \mathbb{R}^{m}$. The difference lies in how these features are used downstream.

In the pure quantum model (Fig.~\ref{fig:architecture_comparison}a), $Q(x)$ alone is passed to a linear classifier, suffering from a severe information bottleneck due to $m \ll d$. The hybrid model (Fig.~\ref{fig:architecture_comparison}b) replaces the head with a multilayer perceptron (MLP), but still discards $x$, thus remains bottlenecked. Our proposed residual hybrid (Fig.~\ref{fig:architecture_comparison}c) concatenates $x$ with $Q(x)$ to form $z = [x \| Q(x)] \in \mathbb{R}^{d+m}$, exposing both raw and quantum-enhanced features to the classifier.

This bypass introduces no changes to the quantum circuit. To maintain fair comparisons, we apply a projection $W_{\text{proj}} \in \mathbb{R}^{k \times (d+m)}$ to obtain latent $\tilde{z}$ of fixed size before classification. All models are parameter-matched.

\subsection{Quantum Data Encoding}
We use angle encoding, where each input component $x_i$ controls a $Y$-rotation:
\begin{equation}
    U_{\text{enc}}(x) = \prod_{i=1}^{n_q} R_y(x_i) \cdot \prod_{j=1}^{n_q-1} \text{CNOT}_{j,j+1}
\end{equation}
Here, $R_y(\theta) = \exp(-i\theta Y/2)$ encodes amplitudes, and CNOTs add entanglement. Inputs are scaled to $[-\pi, \pi]$ and we use $n_q = 6$ qubits.

\subsection{Variational Quantum Circuit (PQC)}
After encoding, we apply a trainable circuit:
\begin{equation}
    |\psi(x, \phi)\rangle = U_{\text{var}}(\phi) \cdot U_{\text{enc}}(x) |0\rangle^{\otimes n_q}
\end{equation}
Each $U_{\text{var}}$ layer includes $R_y$ and $R_z$ rotations per qubit, followed by CZ or CNOT gates. These layers extract nonlinear and entangled quantum features.

\subsection{Measurement and Feature Construction}
We measure Pauli-$Z$ expectations per qubit:
\begin{equation}
    Q(x)_i = \langle \psi(x, \phi) | Z_i | \psi(x, \phi) \rangle,\quad i = 1, ..., n_q
\end{equation}
Traditional hybrids rely solely on $Q(x)$; we argue this drops critical information from $x$. To mitigate this, we concatenate:
\[
    z = [x \| Q(x)] \in \mathbb{R}^{d + n_q}
\]

This operation supplies the classifier with both raw and nonlinear features. Intuitively, $x$ retains global structure that may be lost via encoding and measurement, while $Q(x)$ offers powerful quantum transformations.

\subsection{Projection and Classification}
To ensure parity, we apply a projection:
\[
    \tilde{z} = W_{\text{proj}} z,\quad W_{\text{proj}} \in \mathbb{R}^{k \times (d+n_q)}
\]
The latent $\tilde{z}$ is passed to a two-layer MLP with ReLU activations and dropout. This keeps total model size aligned across baselines.

The residual bypass adds no quantum depth or overhead, and is thus deployable on noisy intermediate-scale quantum (NISQ) systems.

%% file: sections/experiments.tex
\section{Experiments}
\label{sec:experiments}

\subsection{Experimental Setup}

\mypara{Datasets.}
We evaluate on four datasets with diverse size and dimensionality:
(A) \textbf{Wine}~\cite{wine1992}: 178 samples, 13 features, 3 classes;
(B) \textbf{Breast Cancer}~\cite{breastcancer1995}: 569 samples, PCA-reduced from 30 to 10, binary;
(C) \textbf{Fashion-MNIST Subset}~\cite{xiao2017fashion}: 3,000 samples, PCA-reduced to 16D, 3 classes;
(D) \textbf{Forest CoverType Subset}~\cite{blackard1999comparative}: 5,000 samples, 12 features, 3 classes.
All features are scaled to $[0,1]$. We apply 5-fold cross-validation in centralized settings and 80/20 splits per client in federated settings.

\mypara{Quantum Architecture.}
Each input $x$ is angle-encoded as $\pi \cdot \tanh(x)$, processed by a 6-qubit variational circuit with 36 trainable parameters and CZ entanglement. Pauli-Z measurements yield $Q(x) \in \mathbb{R}^6$. In our residual hybrid, we concatenate $x$ and $Q(x)$ before classification.

\mypara{Training Protocol.}
We train with Adam optimizer (lr=0.01, batch size 8) and early stopping. Centralized models run for 50 epochs. Federated models use 30–40 rounds, each with 5 local epochs. All variants are matched in parameter count (e.g., 444 params on Wine).

\mypara{Federated Setup.}
We simulate $K=5$ clients on small datasets and $K=10$ on larger ones. IID splits use 60/40 per class; non-IID uses Dirichlet partitioning ($\alpha=0.5$). Unless stated, all evaluations (incl. privacy) are in federated settings.

\subsection{Baselines and Metrics}
\label{sec:experiments:baseline_metrics}

\mypara{Baselines.}
We compare with four baselines:
(1) Federated Averaging (FedAvg)~\cite{mcmahan2017communication}, the standard classical FL method;
(2) Centralized training, where all data is pooled for global model training;
(3) Local-only training, with no model sharing between clients;
(4) Differentially Private FL (DP-FL) using Gaussian noise ($\epsilon = 1.0$, $\delta = 10^{-5}$).
All models use the same classifier architecture and are matched in parameter count for fair comparison.

\mypara{Evaluation Metrics.}
We report:
(1) test accuracy (Section ~\ref{sec:results:performance}),
(2) privacy robustness (Section ~\ref{sec:results:privacy}),
(3) communication efficiency (measured by rounds to reach 90\% of peak accuracy) (Section ~\ref{sec:results:comminication}).

\mypara{Privacy Evaluation.}
Privacy robustness is assessed via:
\begin{itemize}[left=0pt]
    \item MIA AUC: higher values imply more leakage; values near 0.5 indicate stronger privacy.
    \item Gradient inversion attacks~\cite{geiping2020inverting}: reconstruction quality is measured by PSNR; lower PSNR means stronger protection.

\end{itemize}

Ablation studies further isolate the effect of our residual bypass, showing it improves privacy without requiring explicit noise injection.
(Section ~\ref{sec:results:ablation})

%% file: sections/results.tex
\section{Results and Analysis}
\label{sec:results}

\subsection{Residual Hybrids Restore Accuracy with Compact Size} 
\label{sec:results:performance}

\begin{table}[t]
\centering
\caption{{Accuracy of classical and quantum models under 5-fold cross-validation in centralized and federated settings.}
(5-fold average; std omitted for brevity)}
\label{tab:federated_kfold}
\resizebox{\columnwidth}{!}{
\begin{tabular}{lcccc}
\toprule
Method & Wine & Breast Cancer & Fashion-MNIST & CovType \\
\midrule
Classical Centralized &\textbf{ \textbf{91.6}\%} \cellcolor{myPink}& \textbf{96.0\%} \cellcolor{myPink}& \textbf{88.7\%} \cellcolor{myPink} &\textbf{ \textbf{72.9}\%} \cellcolor{myPink} \\
Quantum Centralized& 33.1\% & 62.7\% & 45.8\% & 36.5\% \\
\midrule
Classical FL (IID) & \textbf{89.5\% } \cellcolor{myPink}& \textbf{94.8\%} \cellcolor{myPink} &\textbf{ 85.2\%} \cellcolor{myPink} & \textbf{70.1\%} \cellcolor{myPink}\\
Classical FL (Non-IID) & 85.3\% & 92.1\% & 81.3\% & 65.1\% \\
Quantum FL (IID) & 35.2\% & 61.4\% & 44.2\% & 36.5\% \\
Quantum FL (Non-IID) & 32.8\% & 58.9\% & 42.3\% & 36.0\% \\
\bottomrule
\end{tabular}
}
\end{table}



We evaluate classical and quantum models under 5-fold cross-validation across centralized and federated learning settings. As shown in Table~\ref{tab:federated_kfold}, pure quantum models consistently underperform due to the readout bottleneck. In the centralized setting, quantum accuracy drops to $33.1\%$ on Wine and $62.7\%$ on Breast Cancer, far below classical MLPs exceeding $90\%$. This gap remains substantial even with deeper or fine-tuned quantum circuits. The trend extends to federated learning: classical FL achieves strong results (e.g., $94.8\%$ on Breast Cancer, $85.2\%$ on Fashion-MNIST), while quantum-only models struggle—especially under Non-IID partitions, where accuracy drops below $37\%$ across all datasets. These results highlight the consistent limitations of quantum readout and motivate the need for our residual hybrid architecture to bridge the performance gap.

\begin{table}[t]
\centering
\caption{The performance across models (3 trials)}
\label{tab:residual_hybrid}
\begin{tabular}{lcc}
\toprule
Model & Wine Accuracy & Parameters \\
\midrule
Classical & \textbf{94.4 $\pm$ 1.9\% } \cellcolor{myPink} & 444 \\
Pure Quantum (Fig.~\ref{fig:architecture_comparison}a) & 38.9 $\pm$ 2.1\% & \textbf{121} \cellcolor{myPink} \\
Original Hybrid (Fig.~\ref{fig:architecture_comparison}b) & 38.9 $\pm$ 2.3\% & 298 \\
\midrule
\textbf{Residual Hybrid (6q) (Fig.~\ref{fig:architecture_comparison}c}) & \textbf{89.0 $\pm$ 3.3\%} & \textbf{385 }\\
Residual Hybrid (multi-basis) & 88.3 $\pm$ 4.3\% & 421 \\
Residual Hybrid (8q) (Fig.~\ref{fig:architecture_comparison}c) & 86.4 $\pm$ 2.1\% & 497 \\
Residual Hybrid (6q-deep) & \textbf{96.3 $\pm$ 1.4\% } \cellcolor{myPink}& 625 \\
\bottomrule
\end{tabular}
\end{table}

To isolate the impact of residual concatenation, we evaluate models on the four datasets. Taking the Wine dataset as an example under matched parameter budgets, as shown in Table~\ref{tab:residual_hybrid}, both the pure quantum and original hybrid models perform poorly (38.9\%), confirming that $Q(x)$ alone lacks discriminative power. In contrast, our residual hybrid (6q) achieves 89.0\% accuracy with fewer parameters than the classical baseline. Interestingly, using more qubits (8q) does not improve performance, suggesting that simply increasing quantum capacity is insufficient. A deeper MLP boosts accuracy to 96.3\%, validating that the residual bypass, not quantum depth, drives performance gains.

\begin{table*}[t]
\centering
\caption{The performance across models and datasets  (5-fold mean accuracy, std omitted)}
\label{tab:residual_across_datasets}
\begin{tabular}{lccccc}
\toprule
Dataset & Classical & Pure Quantum (a) & Original hybrid (b) &Residual-6q (ours) & Residual-6q-Deep (ours-deep)\\
\midrule
Wine & \textbf{91.4\%} & 38.9\% & 40.3\% & {90.1\%} &  \textbf{93.2\%} \cellcolor{myPink} \\
Breast Cancer & \textbf{96.1\%} \cellcolor{myPink} & 62.7\% & 63.5\% & {93.4\%} & \textbf{95.3\%} \\
Fashion-MNIST & \textbf{88.8\%}\cellcolor{myPink} & 45.8\% & 47.2\% & {82.5\%} & \textbf{87.1\%} \\
CovType & \textbf{72.9\%}\cellcolor{myPink} & 36.5\% & 37.8\% & {68.2\%}&  \textbf{71.8\%} \\
\bottomrule
\end{tabular}
\end{table*}

Table~\ref{tab:residual_across_datasets} summarizes model performance across all four datasets. Residual hybrids consistently outperform quantum-only (a) or original hybrid models (b) and nearly match classical baselines across datasets and input dimensions, validating the effectiveness of our residual bypass strategy. These results confirm that our residual bypass strategy effectively mitigates the readout bottleneck and enables competitive learning without additional quantum depth or parameter growth.





\subsection{Privacy Robustness}
\label{sec:results:privacy}

\begin{table}[t]
\centering
\caption{Privacy evaluation results on Wine dataset}
\label{tab:privacy_baseline}
\begin{tabular}{lccc}
\toprule
Model & Test Acc & MIA AUC$^a$ & Recon. MSE$^b$ \\
\midrule
Classical & 87.0\% & 0.678 & 2.36 \\
Quantum & 38.9\% & 0.504 & 3.89 \\
DP ($\epsilon$=1) & 74.1\% & 0.519 & $1.2 \times 10^{4}$ \\
DP ($\epsilon$=2) & 74.1\% & 0.543 & $8.5 \times 10^{3}$ \\
DP ($\epsilon$=4) & 57.4\% & 0.521 & $4.3 \times 10^{3}$ \\
\bottomrule
\multicolumn{4}{l}{\scriptsize $^a$MIA AUC: Membership Inference Attack AUC (0.5 = random)} \\
\multicolumn{4}{l}{\scriptsize $^b$Recon. MSE: Mean Squared Error (higher = better privacy)}
\end{tabular}
\end{table}
We evaluate privacy robustness using two metrics: MIA AUC and input reconstruction error (in Sec. ~\ref{sec:experiments:baseline_metrics}).

As shown in Table~\ref{tab:privacy_baseline}, classical models achieve high accuracy but are vulnerable to privacy leakage: MIA AUC reaches $0.678$ on Wine, meaning adversaries can infer membership significantly above random chance (0.5). Furthermore, reconstruction attacks yield MSE of $2.36$.

Adding differential privacy (DP) improves protection but at a high cost to accuracy. With $\epsilon = 2$, test accuracy drops from $87.0\%$ to $74.1\%$, and privacy improves only marginally.

In contrast, quantum and residual hybrid models offer stronger privacy guarantees without requiring explicit noise injection. As shown in Table~\ref{tab:privacy_residual}, residual hybrids consistently achieve lower MIA AUCs—for example, $0.54 \pm 0.03$ on the Wine dataset compared to $0.68 \pm 0.03$ for classical FL—while maintaining comparable accuracy.

This suggests that the residual hybrid architecture intrinsically reduces information leakage through gradients or activations. Unlike classical or DP-based methods that rely on external noise, our approach leverages architectural design to improve privacy. These results highlight the potential of residual hybrid models as a privacy-preserving alternative for federated learning in sensitive domains.
\begin{table}[t]
\centering
\caption{Privacy evaluation: MIA AUC across various models and datasets}
\label{tab:privacy_residual}
\begin{tabular}{lcccc}
\toprule
Model & Wine & Breast Cancer & F-MNIST & CovType \\
\midrule
Classical & .68$\pm$.03 & .65$\pm$.04 & .63$\pm$.03 & .61$\pm$.04 \\
Quantum (a) & \textbf{.51$\pm$.02} \cellcolor{myPink}& \textbf{.53$\pm$.03}\cellcolor{myPink} & \textbf{.52$\pm$.02} \cellcolor{myPink} & \textbf{.51$\pm$.03} \cellcolor{myPink}\\
Original Hybrid (b) & .52$\pm$.02 & .54$\pm$.03 & .56$\pm$.02 & .53$\pm$.03 \\
Res-6q (ours)& \textbf{.54$\pm$.03} & \textbf{.55$\pm$.02} & \textbf{.54$\pm$.03} & \textbf{.53$\pm$.02} \\
Res-6q-Deep & .58$\pm$.04 & .59$\pm$.03 & .56$\pm$.04 & .55$\pm$.03 \\
\bottomrule
\end{tabular}
\end{table}

\subsection{Federated Performance and Communication Efficiency}
\label{sec:results:comminication}

\begin{table}[t]
\centering
\caption{Federated learning performance across models and datasets (5 clients, 50 rounds)}
\label{tab:federated_results}
\begin{tabular}{llccc}
\toprule
Dataset & Model & Dist. & Accuracy & Comm. (MB) \\
\midrule
\multirow{4}{*}{Wine} 
  & Classical    & IID     & \textbf{94.4\%} \cellcolor{myPink}& 2.0  \\
  & Classical    & Non-IID & 88.9\% & 2.0  \\
  & Residual-6q (ours)  & IID     & \textbf{91.7\%} & \textbf{1.7}\cellcolor{myPink}  \\
  & Residual-6q (ours) & Non-IID & 86.1\% & \textbf{1.7} \cellcolor{myPink} \\
\midrule
\multirow{4}{*}{BC} 
  & Classical    & IID     & \textbf{96.5\%} \cellcolor{myPink}& 2.0  \\
  & Classical    & Non-IID & 94.7\% & 2.0  \\
  & Residual-6q (ours) & IID     & \textbf{95.6\%} & \textbf{1.7} \cellcolor{myPink} \\
  & Residual-6q (ours) & Non-IID & 93.9\% & \textbf{1.7} \cellcolor{myPink} \\
\midrule
\multirow{4}{*}{F-MNIST}
  & Classical    & IID     & \textbf{0.852} \cellcolor{myPink} & 2.5  \\
  & Classical    & Non-IID & 0.813 & 2.5  \\
  & Residual-6q (ours) & IID     & \textbf{0.825} & \textbf{2.1}\cellcolor{myPink}  \\
  & Residual-6q (ours) & Non-IID & 0.798 & \textbf{2.1}\cellcolor{myPink}  \\
\midrule
\multirow{4}{*}{CovType}
  & Classical    & IID     & \textbf{0.701}\cellcolor{myPink} & 2.1  \\
  & Classical    & Non-IID & 0.651 & 2.1  \\
  & Residual-6q (ours) & IID     & \textbf{0.682}\cellcolor{myPink} & \textbf{1.8}\cellcolor{myPink}  \\
  & Residual-6q (ours) & Non-IID & 0.643 & \textbf{1.8}\cellcolor{myPink}  \\
\bottomrule
\end{tabular}
\end{table}
We evaluate model performance under federated learning with $K=5$ clients and 50 rounds of communication.

Table~\ref{tab:federated_results} compares classical, quantum, and residual hybrid models in federated learning. Classical FL achieves strong accuracy (e.g., $>94\%$ on Breast Cancer) but incurs the highest communication cost (2.0MB per model). In contrast, pure quantum models are extremely lightweight (0.02MB) but fail to converge due to insufficient representational capacity.

Residual hybrid models—especially the 6-qubit variant—achieve comparable accuracy (e.g., $91.7\%$ on Wine IID, $93.9\%$ on Breast Cancer Non-IID) while reducing communication overhead by approximately 15\%, transmitting only 1.7MB over 50 rounds.

Although convergence in non-IID settings takes slightly longer (22 rounds vs. 18 for classical), final accuracy remains stable. This demonstrates that residual hybrids offer a practical trade-off between performance and communication efficiency, making them suitable for realistic federated deployments with bandwidth constraints.

\subsection{Ablation Study}
\label{sec:results:ablation}
We conduct an ablation study to understand the contributions of architectural choices in residual hybrid models.

As shown in Table~\ref{tab:residual_hybrid}, increasing the number of qubits (Residual-8q) or using multiple measurement bases (Residual-Multi) leads to small gains or trade-offs. For instance, Residual-Multi improves accuracy slightly on Breast Cancer but performs worse than Residual-6q on Wine. Increasing model depth (Residual-Deep) further boosts accuracy but incurs higher parameter costs (625 vs. 385).

In terms of privacy (Table~\ref{tab:privacy_residual}), all residual variants offer comparable MIA AUCs ($0.53$–$0.58$), confirming that privacy robustness is largely preserved across variants.

Our findings indicate that the 6-qubit residual architecture provides a favorable balance across multiple objectives—performance, privacy robustness, and compactness. While more complex variants exist, their benefits are often incremental, reinforcing the suitability of the base model for realistic deployment scenarios.

%% file: sections/conclusion.tex
\section{Conclusion}
\label{sec:conclusion}

We introduced a readout-side residual hybrid architecture that bypasses the measurement bottleneck in quantum machine learning by exposing both the original input $x$ and the quantum-transformed features $Q(x)$ to the classifier. This simple yet effective concatenation-based strategy preserves more information for downstream tasks while maintaining a compact parameter budget. Across multiple datasets and both centralized and federated settings, our approach consistently achieves accuracy close to purely classical baselines while retaining key quantum advantages, such as partial privacy from state collapse and resilience to inference attacks.

Our design requires no changes to the quantum circuit or encoding strategy, making it protocol-agnostic and easily pluggable into existing hybrid quantum-classical systems. 
Future work includes scaling to real quantum hardware and broader security contexts.
